\documentclass[aps,prl,twocolumn,superscriptaddress,amsmath,10pt]{revtex4}
\usepackage{graphicx}
\usepackage{amsmath}
\usepackage{epsf}
\usepackage{epsfig}
\usepackage{amsmath, amssymb, graphics}
\newcommand{\mathsym}[1]{{}}

\newcommand{\qo}[1]{``#1''}                  
\newcommand{\beqn}{\begin{eqnarray}}         
\newcommand{\eeqn}{\end{eqnarray}}           
\newcommand{\ket}[1]{|#1\rangle}             
\begin{document}
\title{Efficient generation and sorting of orbital angular
momentum eigenmodes of light by thermally tuned q-plates}
\author{Ebrahim Karimi}
\affiliation{Dipartimento di Scienze Fisiche, Universit\`{a} di
Napoli ``Federico II'', Compl.\ Univ.\ di Monte S. Angelo, 80126
Napoli, Italy}
\affiliation{Consiglio Nazionale delle Ricerche-INFM
Coherentia, Napoli, Italy}

\author{Bruno Piccirillo}
\affiliation{Dipartimento di Scienze Fisiche, Universit\`{a} di
Napoli ``Federico II'', Compl.\ Univ.\ di Monte S. Angelo, 80126
Napoli, Italy}
\affiliation{Consorzio Nazionale Interuniversitario per le Scienze
Fisiche della Materia, Napoli}

\author{Eleonora Nagali}
\affiliation{Dipartimento di Fisica, Universit\`{a} di Roma ``La
Sapienza'', 00185 Roma, Italy}

\author{Lorenzo Marrucci}
\affiliation{Dipartimento di Scienze Fisiche, Universit\`{a} di
Napoli ``Federico II'', Compl.\ Univ.\ di Monte S. Angelo, 80126
Napoli, Italy}
\affiliation{Consiglio Nazionale delle Ricerche-INFM Coherentia, Napoli,
Italy}

\author{Enrico Santamato}
\affiliation{Dipartimento di Scienze Fisiche, Universit\`{a} di
Napoli ``Federico II'', Compl.\ Univ.\ di Monte S. Angelo, 80126
Napoli, Italy}
\affiliation{Consorzio Nazionale Interuniversitario per le Scienze
Fisiche della Materia, Napoli}
\email{enrico.santamato@na.infn.it}

\begin{abstract}
We present methods for generating and for sorting specific orbital
angular momentum (OAM) eigenmodes of a light beam with high
efficiency, using a liquid crystal birefringent plate with unit
topological charge, known as \qo{q-plate}. The generation efficiency
has been optimized by tuning the optical retardation of the q-plate
with temperature. The measured OAM $m=\pm2$ eigenmodes generation
efficiency from an input TEM$_{00}$ beam was of 97\%. Mode sorting
of the two input OAM $m=\pm2$ eigenmodes was achieved with an
efficiency of 81\% and an extinction-ratio (or cross-talk) larger
than 4.5:1.
\end{abstract}
\maketitle
\begin{figure}[htb]
\centerline{\includegraphics[width=5cm]{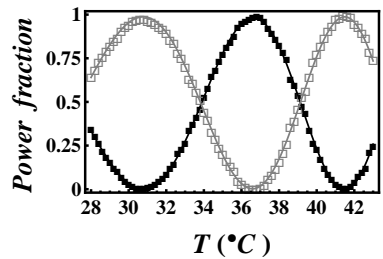}}
\caption{\label{fig:2} STOC power fraction $P_{R,2}/P_0$ (black
squares) and no STOC power fraction $P_{L,0}/P_0$ (empty squares) as
functions of the QP temperature. The curve is the best fit obtained
as explained in the text.}
\end{figure}

\noindent A light beam has two \qo{rotational} degrees of freedom:
spin angular momentum (SAM) and orbital angular momentum (OAM).
Light SAM is related to the vectorial properties of the transverse
electric field and may take two values $s=+1$ and $s=-1$ (in units
of $\hbar$), corresponding to left and right circularly polarized
light, respectively. Light OAM is defined by the phase structure of
the complex electric field~\cite{allen} and may take any of the
infinite values $m=0,\pm 1,\pm 2,\dots$. In the last decade, the
interest in light beams endowed with OAM has continuously increased,
because of the wide range of scientific and technological
applications in both classic~\cite{gibson04} and quantum regimes of
light~\cite{mair01,molinaterriza07}. Till today,
only few tools have been developed for generating and manipulating
OAM, including pitch-fork holograms~\cite{bazhenov90}, spiral phase
plates~\cite{beijersbergen94,sueda04}, Dove prisms, Leach's
interferometers~\cite{leach04a}, and cylindrical lens mode
converters~\cite{allen92}. All these devices and techniques have
drawbacks and limitations in terms of efficiency, modulation speed,
working wavelength, alignment, and constraints imposed on the input
and output beams.\\
Recently, a novel optical device for the OAM manipulation has been
introduced, made as a birefringent liquid crystal plate having a
azimuthal distribution of the local optical axis in the transverse
plane, with a topological charge $q$ at its center defect, hence
named \qo{q-plate} (QP). When a light beam traverses a QP, the
topological charge $2q$ is transferred into the phase of the beam,
which thus gains a corresponding amount of OAM, with a sign
determined by the light input polarization. In this paper we
consider only QP with $q=1$, having cylindrical symmetry around
their central defect. Because of this symmetry, a $q=1$ QP cannot
change the total SAM + OAM angular momentum of the incident beam, so
that its action is just that of converting the SAM-variation of some
photons into OAM, and \textit{vice versa} (SAM-to-OAM-conversion, or
STOC)~\cite{marrucci06}. Besides the topological charge $q$, the QP
is characterized by its birefringent retardation $\delta$, ideally
uniform in the transverse plane, which determines the STOC
efficiency, i.e. the fraction of photons (or optical energy) that is
actually converted. Moreover, QPs are highly transparent and can be
cascaded along the beam to produce arbitrary values (even, if $q=1$)
of the OAM~\cite{marrucci06a}. The STOC efficiency can ideally touch
100\%~\cite{karimi09}. In this work we achieved STOC efficiencies
exceeding 95\%, to be compared, for example, with the record
efficiencies ranging from 50\% to 90\%, depending on the wavelength,
of computer-generated blazed fork
holograms~\cite{heckenberg92a,he95}. High efficiencies in producing
and detecting the light OAM are highly desirable in all
circumstances where only few photons are at disposal. Examples are
weak signals coming from far sources, such as astronomical
ones,~\cite{harwit03} or after propagation in highly absorbing
media, or in quantum information applications. Achieving high
efficiencies, however, requires accurate tuning of the QP
retardation $\delta$. In order to tune the optical retardation
$\delta$ and thus optimize the QP efficiency, in this work we
adopted a method based on controlling the material temperature,
which presents good features in terms of a realization simplicity
and stability of the obtained retardation.\newline
This paper is divided in two parts, the first dealing with OAM
generation and the second with OAM sorting, e.g., for detection
purposes. In the first experiment, we used the STOC process for
transforming an input TEM$_{00}$ laser beam into a beam having OAM
$m=\pm2$. Our QP was manufactured as a $6 \mu$m thick film of E7
liquid crystal from Merck, sandwiched between two circularly rubbed
glass substrates coated with the polyimide PI2555 from DuPont. The
transmittance $T$ of our QP was measured to be 88\%, with the losses
arising from scattering due to manufacturing imperfections and from
the lack of antireflection coating on the cell bounding glasses.
\begin{figure}[htb]
\centerline{\includegraphics[width=7.8cm]{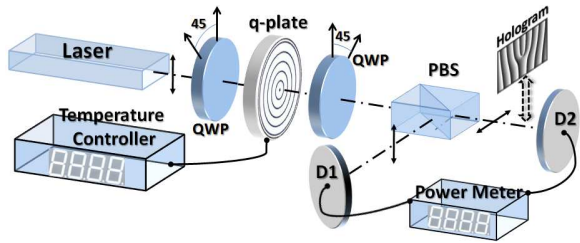}}
\caption{\label{fig:1} Setup to measure the STOC efficiency and the
state purity. Legend: QWP - quarter-wave plate; PBS - polarizing
beam-splitter. The fork hologram was inserted on the converted beam
arm for verifying the degree of purity of the OAM $m=2$ mode
generated on the output.}
\end{figure}
The optical setup used to measure the STOC efficiency of our QP as a
function of its optical retardation is shown in Fig.~\ref{fig:1}.
The input light was a linearly polarized TEM$_{00}$ laser beam
generated by a frequency-doubled continuous-wave Nd:YVO$_4$
($\lambda=532$~nm). After changing the polarization into
left-circular ($L$) by a suitably oriented quarter wave plate (QWP),
the beam was made to pass through our controlled-temperature QP. In
the QP, a fraction of the photons will undergo the STOC process and
will therefore emerge with right-circular ($R$) polarization and OAM
$m=2$, and the others will remain in the OAM $m=0$ and with $L$
polarization. For arbitrary QP tuning, the transverse intensity
pattern of the beam emerging from the QP exhibits a central spot,
corresponding to the light fraction that is left in the initial OAM
state $m=0$, surrounded by a single ring, corresponding to the light
converted into the OAM $m=2$ mode (doughnut beam). By inserting in
the output beam a second QWP and a polarizing beam-splitter (PBS)
oriented so as to select the $R$-polarization for, say, the
transmission output, a pure doughnut beam is obtained. The reflected
output of the PBS shows instead only the central spot (unconverted
light). If $P_{in}$ is the total input power, the respective powers
of the coherently converted and unconverted components, $P_{R,2}$
and $P_{L,0}$, are expected to depend on the optical retardation
$\delta$ according to the following Malus-like
laws~\cite{marrucci08,karimi09}:
\begin{equation}\label{eq:STOC}
    P_{R,2} = P_0\sin^2\frac{\delta}{2}\hspace{0.5cm}
    P_{L,0} = P_0\cos^2\frac{\delta}{2}
\end{equation}
where $P_0 = T P_{in}$ is the total power transmitted coherently by
the QP. To adjust the retardation $\delta$, the temperature of the
QP was varied while measuring the power of the two output beams of
the PBS. The results are shown in Fig.~(\ref{fig:2}), together with
best-fit curves based on Eqs.~(\ref{eq:STOC}), assuming a
second-order polynomial dependence $\delta(T)=a+bT+cT^2$ and adding
a constant offset of 0.5\% that accounts for the finite PBS and
wave-plates contrast ratios. When the PBS-transmitted power (full
squares in Fig.~(\ref{fig:2})) reaches its maximum, we obtain the
optimal STOC and almost all photons emerge in the $m=2$ OAM state.
More precisely, in this optimal situation, about 99.2\% of the beam
power is transmitted by the PBS, and after taking into account the
finite contrast ratio of the waveplates and PBS (as measured without
the QP), the actual QP efficiency in inverting the optical
polarization is estimated to be 99.6\%. Near the minima of the same
curve, the STOC process is minimum and almost all photons emerge in
the original $m=0$ state.
\newline
To test the purity of the OAM eigenmode generated by our QP, at the
optimal temperature we inserted along the beam a double pitchfork
hologram as OAM-mode splitter~\cite{bazhenov90,heckenberg92a} and,
on the first-order diffracted beam we selected the central spot by a
suitable iris placed before the detector. After suitable calibration
of the detection efficiency, the measured OAM $m=2$ mode content
fraction was estimated to be $F=97.2\%$ (in quantum optics, $F$ is
the squared ``fidelity'' to the desired mode $m=2$), so that the
overall QP efficiency in generating a pure OAM $m=2$ mode is $\eta =
97.2\%\times99.6\% = 96.9\%$ (this value is net of reflection and
scattering losses in the QP; including all losses, the efficiency of
our QP is 85\%, a figure which could be however easily improved to
more than 90\% by simply adding antireflection coatings). Moreover,
we note that in order to invert the sign of the generated OAM value
$m$ (e.g., from $+2$ to $-2$) with our setup it is enough to switch
the input and output polarizations to the orthogonal ones. This can
be achieved at gigaHertz rates by means of Pockel cells. No
switching capability is instead possible with passive holograms,
while computer-controlled spatial light modulators (SLM) can achieve
at most switching rates of the order of few kiloHertz.
\newline
Let us now discuss the second experiment, about OAM mode sorting.
More precisely, we present a setup for sorting the four modes given
by combining the two OAM modes $m=2$ and $m=-2$ and the two
orthogonal polarizations $L$ and $R$. The setup is similar to the
previous one, except that the input laser beam was made to pass
through a SLM driven with a computer generated hologram (CGH) for
determining its input OAM state. The temperature of the QP was held
fixed at the optimal value for maximum STOC efficiency. We used
double-pitchfork CGHs to produce alternatively $m=2$ and $m=-2$ OAM
eigenstates in the input. The first QWP was also rotated so as to
produce, alternatively, right-circular and left-circular
polarizations.
\begin{figure}[h]
\centerline{\includegraphics[width=8.8cm]{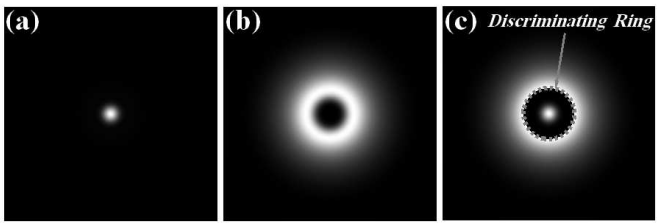}}
\caption{\label{fig:3} Calculated far-field patterns of OAM modes
$m=0$ and $m=4$, generated by the q-plate for input OAM $m=\pm2$
(The input beam was assumed
HyGG$_{-2,\pm2}(r,\phi,0.1)$~\cite{karimi07} mode). The dashed
circle shows the discriminating area used in the balanced mode
sorter.}
\end{figure}
In this way, we created in sequence the four photon states
$\ket{L,2}$, $\ket{L,-2}$, $\ket{R,2}$, $\ket{R,-2}$, where the
first symbol in the ket denotes the polarization and the second is
the $m$ value of the photon OAM. Because the STOC process is
complete in a tuned QP, after passing through the QP these four
states are expected to change respectively into $\ket{R,4}$,
$\ket{R,0}$, $\ket{L,0}$, $\ket{L,-4}$. The QWP after the QP,
changes these states into $\ket{H,4}$, $\ket{H,0}$, $\ket{V,0}$,
$\ket{V,-4}$, respectively, so that the two states $\ket{H,4}$,
$\ket{H,0}$ are transmitted by the PBS and the other two states
$\ket{V,0}$, $\ket{V,-4}$ are reflected. We see that owing to the
QP, the two states in each of the reflected and transmitted beam
have a different value of photon OAM ($m=0$ and $m=4$). After
propagating in the far-field (or in the focal plane of a lens),
these two modes can then be separated by exploiting their different
radial distribution, i.e., a central spot for $m=0$ and an outer
ring for $m=4$, as shown in Fig.~(\ref{fig:3}), thus finally sorting
all four initial spin-orbit modes into separate beams. The radial
sorting can be obtained, for example, by means of a mirror with a
hole at its center. We note that in our measurements we are using
the PBS only for discriminating the two input polarizations. In many
applications, the input polarization state is fixed and known and
one is interested only in sorting the OAM $m=\pm2$ modes. In this
case, the PBS is unnecessary, the input light can be always made
$L$-polarized, by suitable wave plates, and the output beam will
then be only $R$-polarized. The OAM sorting is then still based on
the radial-mode separation in the far field. The efficiency of this
mode-sorter, here defined as the fraction of the optical power in a
given eigenmode to be sorted that is directed in the \emph{correct}
output mode is however not 100\%, because of the radial mode
overlap, leading to some energy going in the ``wrong'' OAM output
mode. This also leads also to a finite contrast ratio, i.e., to
cross-talk between the input channels. In Table~\ref{tab:tab1} we
report the measured efficiencies and contrast ratios for the four
input spin-orbit base states previously mentioned, with a
discriminating hole radius chosen so as to balance the output
efficiencies for opposite input OAM (see Fig.~(\ref{fig:3})). The
measured efficiency of the QP as mode sorter is of about 81.5\%
(72\% with QP losses), i.e., 2.7 times larger than for the best
available holograms ($\approx 30$\% efficiency, as blazing cannot be
used for multiple outputs). The extinction ratio due to radial
overlap between $m=0$ and $m=4$ OAM modes can be improved by
introducing a suitable opaque belt mask that cuts away the
overlapping region of the two modes, although at the expense of a
reduced efficiency. In principle the contrast ratio can be made
arbitrarily large. Theoretically, we estimate a contrast ratio
$>10^3$ for an efficiency of about 50\% and $>10^6$ for an
efficiency of 10\%. We note that the radial-overlap problem leading
to cross-talk is not unique of the QP approach; a similar problem
and an equivalent efficiency/contrast-ratio tradeoff is present also
with hologram-based OAM sorting. We also tested our QP mode-sorter
with coherent superpositions of $m=+2$ and $m=-2$ OAM modes
(obtained with suitable CGHs), obtaining results consistent with the
efficiencies reported in Table~\ref{tab:tab1}.\newline
\begin{table}[h]
  \centering
  \caption{\label{tab:tab1} The QP's efficiency as a mode sorter.}
  \begin{tabular*}{0.43\textwidth}{@{\extracolsep{\fill}} cccc}
  \hline\hline
  Input state  & Output state & Efficiency & Extinction ratio \\ \hline
  $\ket{L,2}$  & $\ket{R,4}$  & 81.1\%   & $\approx$ 4.6:1 \\
  $\ket{L,-2}$ & $\ket{R,0}$  & 81.8\%   & $\approx$ 4.5:1 \\
  $\ket{R,2}$  & $\ket{L,0}$  & 81.6\%   & $\approx$ 4.7:1 \\
  $\ket{R,-2}$ & $\ket{L,-4}$ & 81.5\%   & $\approx$ 4.6:1 \\
  \hline\hline
  \end{tabular*}
\end{table}
In conclusion, we demonstrated the application of a liquid crystal
q-plate as (i) a switchable OAM generator and (ii) a OAM
mode-sorter, exploiting the optical spin-to-orbital angular momentum
conversion process. By suitable thermal tuning of the birefringent
retardation in the q-plate, we optimized its generation or
mode-sorting efficiency, finding values as large as $\approx$ 97\%
as mode generator and $\approx$ 81\% as mode sorter. These results
show that the q-plates can easily reach much larger efficiencies
than holographic elements and will therefore provide the most
convenient option for many applications, particularly when low
photon fluxes are available.

\end{document}